# Surface Reactivity Enhancement by $O_2$ Dissociation on a Single-layer MgO Film Deposited on Metal Substrate

Cequn Li[†], Jing Fan[†], Bin Xu[§], and Hu Xu[*,†]

[†]Department of Physics, South University of Science and Technology of China, Shenzhen, 518055, China

[§]Physics Department and Institute for Nanoscience and Engineering, University of Arkansas, Fayetteville, Arkansas 72701, USA

**ABSTRACT:** Improving reactivity on an insulating surface is crucial due to their important applications in surface catalytic reactions. In this work, we carried out first-principles calculations to investigate the adsorption of $O_2$ on a single-layer MgO(100) film deposited on metal substrate. The adsorption configurations, reaction pathways, molecular dynamics simulations, and electronic properties are reported. We reveal that $O_2$ can completely dissociate on the surface, which is in sharp contrast to that on MgO(100) films thicker than one monolayer. The dissociated $O_2$ tends to penetrate into the interfacial region, behaving like a switch to trigger subsequent chemical reactions. As an example, the interplay between water and the interfacial oxygen results in the formation of hydroxyl radicals. This study paves an avenue to accomplish the desired surface catalytic reactions, especially those involving oxygen.

## ■ INTRODUCTION

The interaction of $O_2$ with metal oxide surfaces has attracted extensive attention because of their important applications in chemical sensing, catalytic reaction, and photocatalysis.[1,2] Understanding how to activate $O_2$, especially to split oxygen molecules on metal oxide surfaces at the molecular level is of fundamental interests to uncover the catalytic oxidation reaction.[3,4] Usually, molecular oxygen binds weakly on a stoichiometric metal oxide surface, while it can strongly interact with the reduced oxide surfaces in the presence of oxygen vacancies or cation interstitials by forming either superoxo $(O_2^-)$[5] or peroxo $(O_2^{2-})$[6] depending on the extent of charge transfer.

MgO(100) is one of the typical insulating surfaces, and it is usually inactive in surface reactions. For the past few years, metal-supported ultrathin MgO films have been extensively studied[7-14] due to their promising applications in surface chemistry and photocatalysis. The interaction between insulating oxide films with metal substrates may result in the enhancement of chemical reactivity.[12-14] For instance, $O_2$ is predicted to be activated on an ultrathin MgO(100) film supported on Ag(100) with an adsorption energy of -0.82 eV.[13] Subsequently, the formation of $O_2^-$ radical on the

Mo(100) supported MgO(100) ultrathin films is confirmed by electron paramagnetic resonance spectroscopy.[14] The charge transfer due to the tunneling mechanism between $O_2$ and the underlying substrate is responsible for the activation of $O_2$.[13] The activated $O_2$ is an important intermediate in surface catalytic reactions.[13-16] Usually, ultrathin MgO(100) films are prepared by deposition of Mg in the presence of oxygen, i.e., by oxidation of Mg.[17] In this situation, extra oxygen atoms can accumulate at the MgO/Ag interface at high growth temperature.[17-18] As the interfacial oxygen atoms spontaneously oxidize Ni nanoclusters[18], they may also participate in many other chemical reactions. Therefore, it is very interesting to uncover the dissociation mechanism and to know how to accomplish the complete dissociation of $O_2$ on MgO(100) surface.

In this work, we have investigated the adsorption of $O_2$ on one monolayer (ML) MgO(100) deposited on Ag(100) substrate using first-principles calculations. The results reveal that the adsorption behavior of $O_2$ on 1 ML MgO(100)/Ag(100) is essentially different from that on thick (n≥2) MgO(100) films. The adsorbed $O_2$ is energetically favorable to dissociate and tend to locate at interfacial region on 1 ML MgO(100)/Ag(100), while it is chemically adsorbed on thick (n≥2) MgO(100) films in the molecular form. We further demonstrate that the dissociation of $O_2$ has potential applications in surface reactions. For instance, our results clearly show that water molecules split due to the interaction with the dissociated O atoms to form OH radicals, suggesting that 1 ML MgO(100)/Ag(100) surface plays an important role in surface catalytic reactions as a good catalyst.

## ■ METHODS

The first-principles calculations are performed using the Vienna Ab-initio Simulation Package (VASP).[19-21] Projector augmented-wave (PAW) method[22-24] and Perdew-Burke-Ernzerhof (PBE) exchange correlation functional[25-26] are used. The energy cutoff is 450 eV, and each atom is relaxed until the Hellmann-Feynman force is less than 0.02 eV/Å. Neighboring slabs are separated by a vacuum of 12 Å to avoid image interactions. The (2×2×1) and (4×4×1) $k$-point samplings are employed for structural relaxations and electronic properties, respectively. The climbing image nudged elastic band (CI-NEB) method[27] is used to investigate the reaction pathway and barriers. A p(4×4) slab of four Ag layers is adopted to mimic the substrate, which is found to be converged with respect to the number of layers. The Bader charge analysis[28-29] is used to calculate the charge transfer. To study the dynamical behavior of $O_2$ dissociation, we carry out *ab initio* molecular dynamics simulations.

## ■ RESULTS AND DISCUSSION

$O_2$ can only physisorb on the stoichiometric MgO(100) surface, while $O_2$ interacts strongly with MgO(100) ultrathin films deposited on Ag(100)[13] or Mo(100)[14] substrates. For the adsorption of $O_2$ on 1 ML MgO(100)/Ag(100) surface, two adsorption configurations are possible, as shown in Figures 1(a) and 1(b). In the bridge configuration (Figure 1(a)), $O_2$ binds to two surface Mg, and the calculated adsorption energy per $O_2$ is -0.70 eV. This configuration was found to be favorable for $O_2$ adsorption on 2 ML or other multi-layered MgO(100)/Ag(100) ultrathin films.[13] The other adsorption configuration is shown in Figure 1(b), in which $O_2$ tends to locate on top of the surface lattice O with its molecular axes parallel to one chain of the surface oxygen (named as top configuration). For top configuration, each oxygen of $O_2$ binds to two surface Mg, and the calculated bond length is 2.12 Å. Pronounced surface relaxation is caused by $O_2$ adsorption, and the underlying O is pushed downward significantly. Interestingly, $O_2$ in top configuration has much lower adsorption energy (-0.84 eV per $O_2$) than that in bridge configuration, which suggests that bridge configuration becomes energetically less favorable for $O_2$ adsorption with the thickness of MgO(100) down to 1 ML. Due to the strong interaction between the adsorbed $O_2$ and metal substrate, the interfacial bond length of O-Ag decreases from 2.62 Å in the bridge configuration to 2.11 Å in the top configuration. Furthermore, the bond length of $O_2$ in top configuration becomes longer by 0.26 Å upon adsorption. In the following, only the top configuration will be considered.

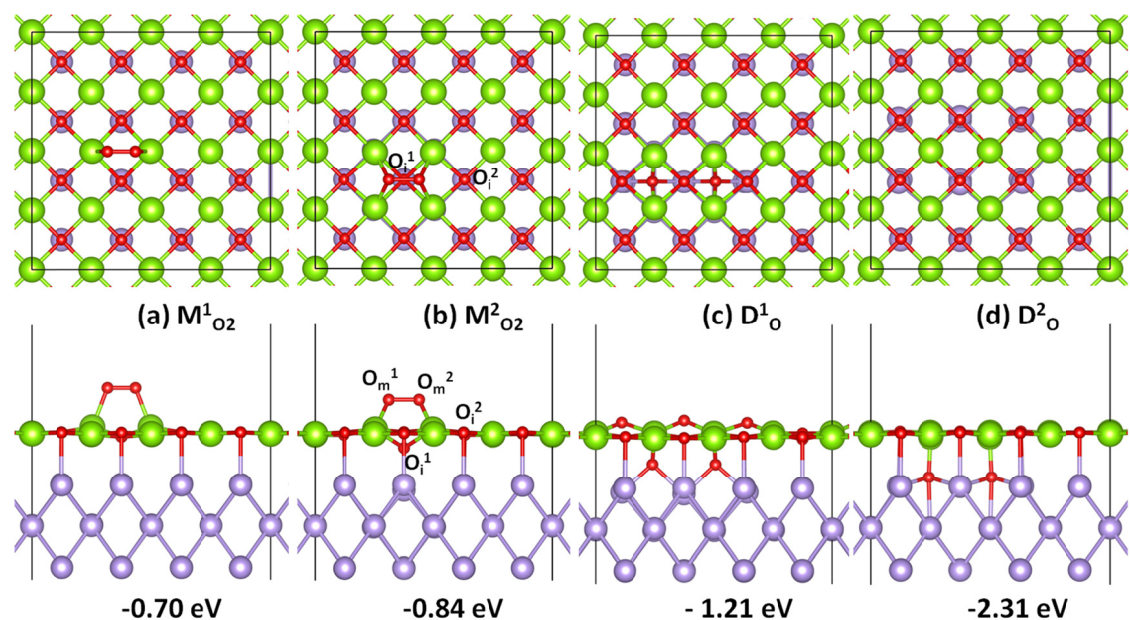

**Figure 1.** Top and side views of adsorption structures of $O_2$ on 1 ML MgO(100)/Ag(100). (a) $O_2$ binds to two surface Mg in molecular form, and (b) molecular $O_2$ binds to four surface Mg. $O_2$ adsorption in dissociative form located (c) at the hollow sites and (d) beneath Mg cations in the interfacial region.

To study the possible dissociative adsorption of $O_2$ on 1 ML MgO(100)/Ag(100) surface, two typical dissociation configurations of $O_2$ are considered, i.e., the hollow sites in the interfacial region and under the surface Mg, as shown in Figures 1(c) and 1(d), respectively. The adsorption energies per $O_2$ are, respectively, -1.21 eV and -2.31 eV for the hollow and beneath cases. To our surprise, $O_2$ adsorption on 1 ML MgO(100)/Ag(100) in the dissociative form has much lower adsorption energy than that in the molecular form, which is in sharp contrast to the case on 2 or more ML MgO(100)/Ag(100) surfaces. The bond length between the dissociated O and surface Mg is 2.01 Å. The interfacial sites just below the surface Mg will be considered next, as this configuration is energetically more favorable for the dissociated $O_2$.

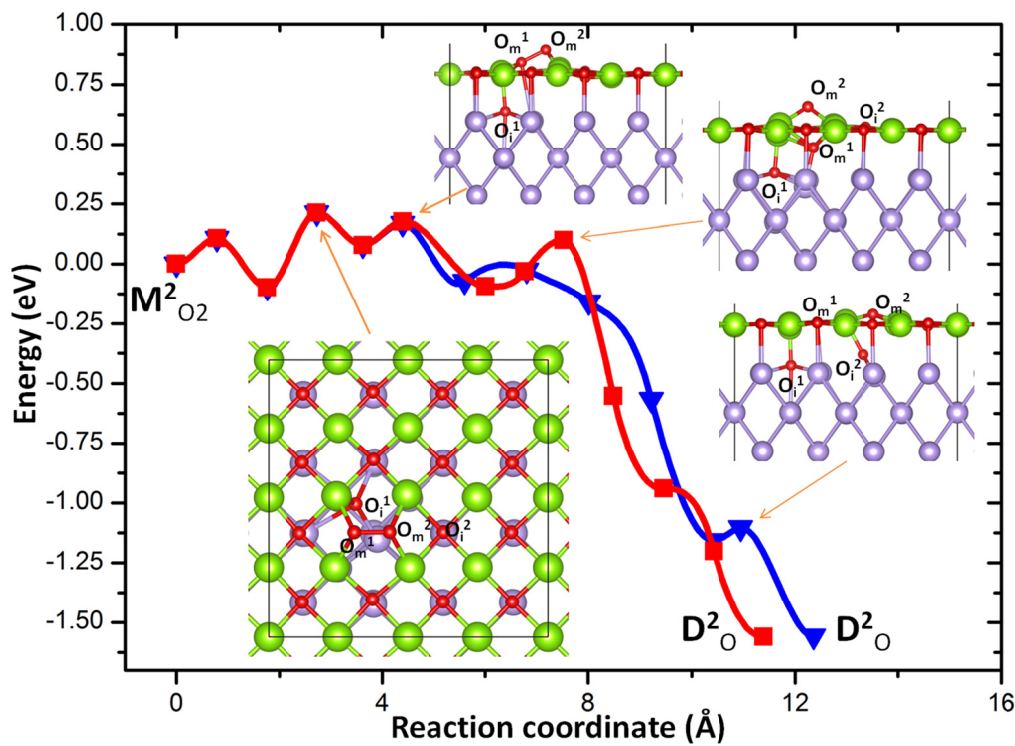

**Figure 2.** Two possible potential energy profiles for $O_2$ dissociation on 1 ML MgO(100)/Ag(100) surface. The final state $D^2_O$ is shown in Figure 1(d).

Now let us turn to the kinetics from adsorption to dissociation. Two possible reaction pathways for $O_2$ dissociation on 1 ML MgO(100)/Ag(100) surface are shown in Figure 2. The reaction barriers in splitting $O_2$ into two individual interfacial O atoms are relatively small, which is about 0.3 eV. By contrast, the energy barrier for $O_2$ dissociation on clean Ag(100) is around 1.0 eV.[30] Therefore, it is more likely for $O_2$ to dissociate on 1 ML MgO(100)/Ag(100) surface, indicating the strong substrate-induced activation of $O_2$. For the dissociation pathway indicated by the blue line, the adsorbed $O_2$ first push the surface $O_i^1$ into the interface region under the surface Mg cation by climbing over a small barrier of 0.3 eV; then the O vacancy is filled with $O_m^1$ of $O_2$. Afterwards, $O_m^2$ from the dissociated $O_2$ can easily stride over surface Mg-Mg bridge and push $O_i^2$ into the interface region. Another reaction pathway is indicated by the red line, for which $O_m^1$ fills the O vacancy when $O_i^1$ is pushed into the interface region and then diffuse to the interfacial region with a small barrier of 0.2 eV. Finally, $O_m^2$ heals the same O vacancy. To further confirm the dissociation of $O_2$ on 1 ML MgO(100)/Ag(100) surface, we also performed molecular dynamics simulations at 300 K, as shown in Figure 3. It demonstrates that $O_i^1$ is pushed downwards at around 7 ps, while $O_i^2$ diffuses into the interfacial region at around 33 ps. Our molecular dynamics simulations at room temperatures also indicate that $O_2$ prefers to dissociate on 1 ML MgO(100)/Ag(100) surface.

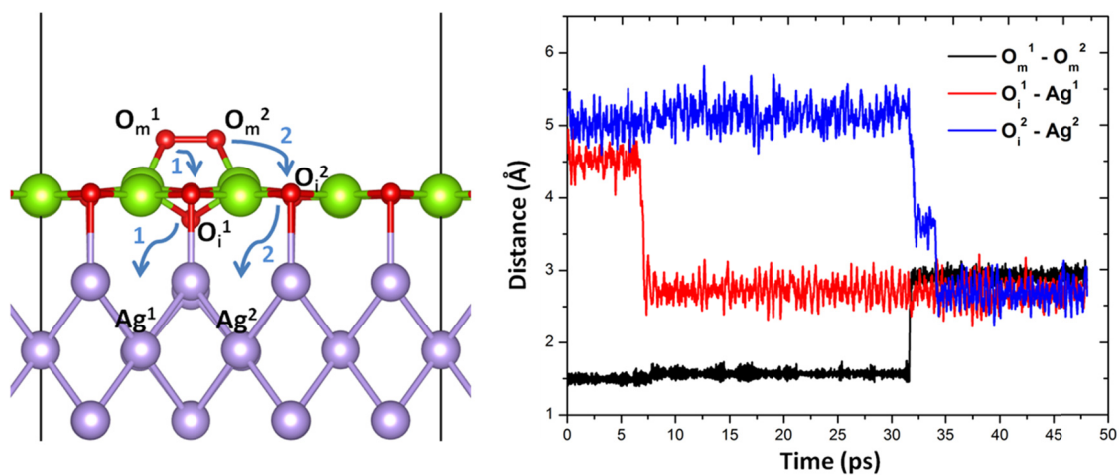

**Figure 3.** The molecular dynamics simulations of $O_2$ dissociation on 1 ML MgO(100)/Ag(100) surface at 300 K. The bond lengths of $O_m^1$-$O_m^2$, $O_i^1$-$Ag^1$, and $O_i^2$-$Ag^2$ are labelled by black, red and blue lines, respectively.

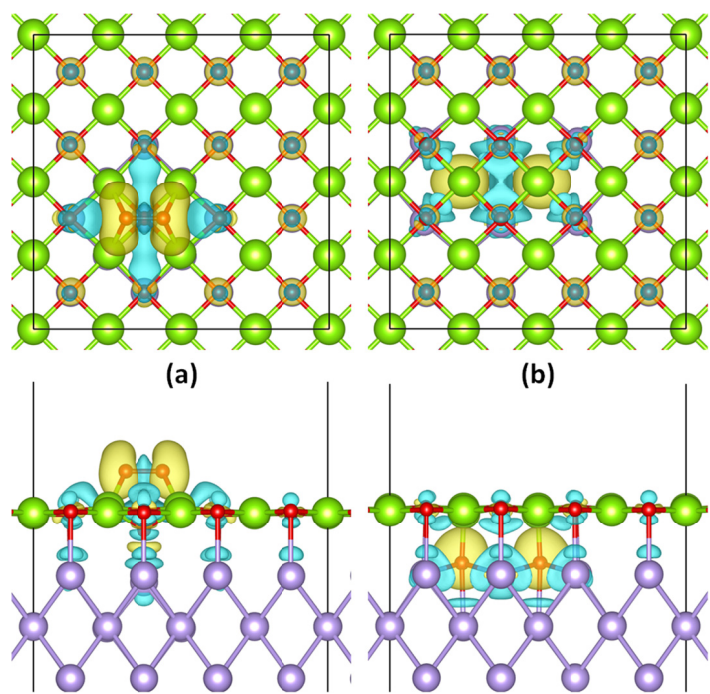

**Figure 4.** Top and side views of differential charge densities for (a) molecular and (b) dissociative adsorption of $O_2$ on 1ML MgO(100)/Ag(100) surface. The used isosurface values are 0.003 e/bohr$^3$ (yellow: charge accumulation, ice-blue: charge depletion).

In the following, let us look at the electronic properties of the adsorbed $O_2$, viz., the Bader charges, density of states, and differential charge densities. It is well known that the charge transfer between the adsorbed $O_2$ and MgO(100) surface deposited on Ag(100) substrate is due to the substantial polaronic distortion.[31,32] As shown in Figure 4(a), the adsorbed $O_2$ in top configuration is negatively charged due to work function reduction of the Ag(100), and $O_2$ with high electron affinity prefers to gain electrons (yellow regions) from the underlying 1 ML MgO film, resulting in strong ionic bonds between oxygen and the 1 ML MgO film. Meanwhile, the charge density is depleted around the adsorption site nearby (ice-blue regions). In addition, the Bader analysis shows that the charge gains per $O_2$ on 1 ML MgO(100)/Ag(100) is 1.52 e, indicating the formation of $O_2^{2-}$ on the surface. In contrast, the amount of charge transfer between $O_2$ and 2 ML MgO(100)/Ag(100) is only 0.72 e.[13] After $O_2$ dissociation, the dissociated O atoms go to the interfacial region. From the differential charge density shown in Figure 4(b), the amount of charge accumulation and depletion around the dissociated O atoms is significant, indicating the strong bonding character. Bader analysis shows that each dissociated O atom gains 1.11 e from the 1 ML MgO/Ag system. In addition, we also analyze the density of states as shown in Figure 5. In Figure 5(a), it clearly shows that 1 ML MgO(100) surface has gap states induced by Ag substrate due to their interfacial chemical bonding.[32] We can find two discrete peaks for the adsorbed $O_2$ near the Fermi level shown in Figure 5(b). After the splitting of $O_2$, the electronic states for the interfacial oxygen are shifted to the lower energy levels, as shown in Figure 5(c). The electronic states of the interfacial oxygen will be slightly shifted to the lower energy region by formation OH groups shown in Figure 5(d), indicating the strong chemical bonding between OH and 1 ML MgO(100)/Ag(100) surface.

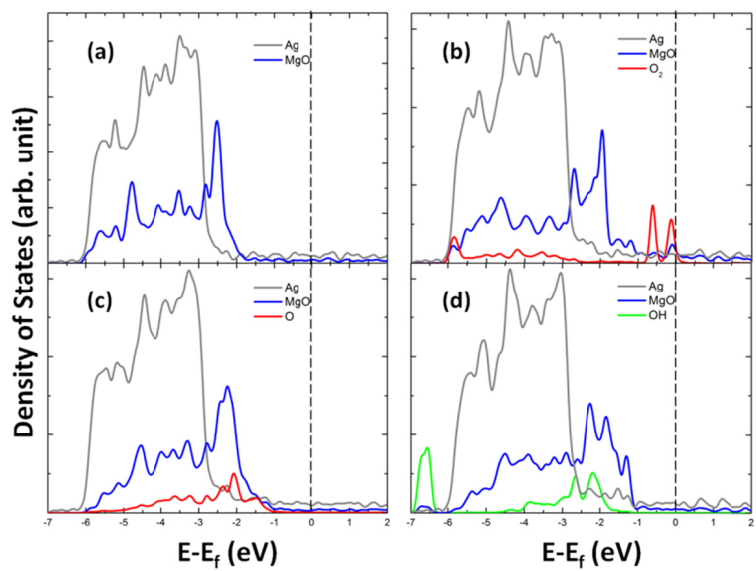

**Figure 5.** Density of states of 1 ML MgO(100)/Ag(100) surface. (a) Clean surface, (b) the adsorbed $O_2$, (c) the dissociated $O_2$, and (d) fully dissociated OH groups. The Fermi levels are shifted to zero.

As mentioned above, $O_2$ can dissociate on 1 ML MgO(100)/Ag(100) surface, and the dissociation in the interfacial region may play a crucial role in surface reactions. Water is frequently used in chemical reactions as solvent, so it is interesting to study the interaction between water and the dissociated $O_2$. It is well known that water prefers to molecularly adsorbed on the stoichiometric MgO(100) surface at the low coverage[33]. In addition, water adsorption on MgO(100)/Ag(100) has been widely studied by Kawai *et. al.*[8-10]. Recently, the substrate-induced strain is proposed to play a crucial role in controlling the dissociation of single water molecule on ultrathin oxide films[34]. In the presence of interfacial O, water prefers to be adsorbed above the surface Mg, as shown in Figure 6(a), and there is one strong hydrogen bond between water and surface oxygen with a distance of 1.61 Å. The adsorption energy per water is -0.61 eV, which is slightly lower than that (-0.50 eV) of on 1 ML MgO(100)/Ag(100) in the absence of interfacial O. Furthermore, the adsorption energy of dissociated $H_2O$ (Figure 6(b)) is -0.66 eV, implying that dissociation of the water molecule on the 1 ML MgO(100)/Ag(100) surface is slightly favorable. In other words, the interfacial oxygen atoms will results in water dissociation on 1 ML MgO(100)/Ag(100) surface.

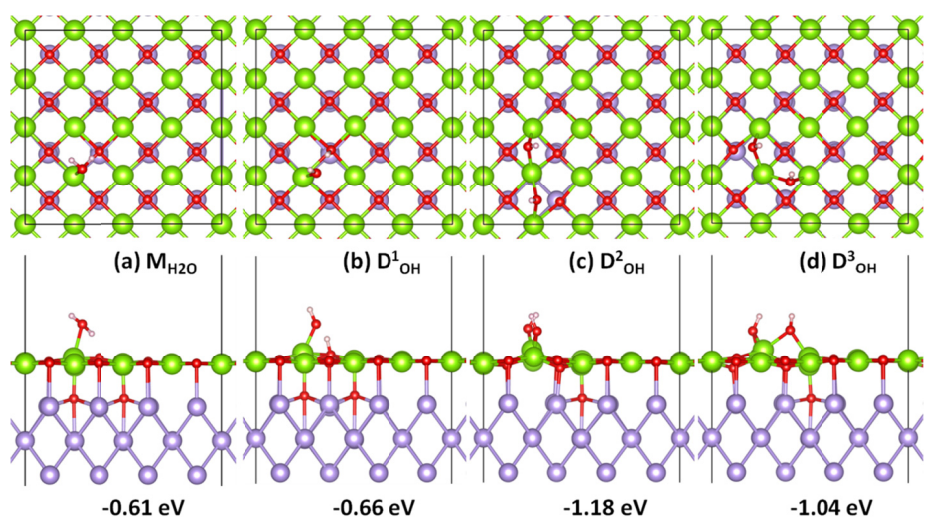

**Figure 6.** Top and side views of water adsorption configurations on 1 ML MgO(100)/Ag(100) surface in the presence of the interfacial oxygen atoms. Water adsorbs in (a) molecular and (b) dissociative forms. (c-d) The dissociated water reacts with the interfacial oxygen to form another OH radical.

The interaction of water with the interfacial O may also leads to the formation of additional OH groups. If one interfacial O is pulled out, water interacts strongly with this O by forming two OH groups. Two possible dissociation configurations are shown in Figures. 6(c) and 6(d), and the corresponding adsorption energies per water are -1.18 eV and -1.04 eV, respectively, indicating that the structure shown in Figure 6(c) is more stable, and the total energy is further reduced by 0.47 eV in the process of water dissociation compared with that in Figure 6(a). The calculated reaction pathway is shown in Figure 7. First, water dissociates via transferring an H to one surface O nearby by overcoming a small energy barrier of 0.03 eV. Then, the interfacial O atom below the Mg cation site is pulled out with a small associated barrier of 0.4 eV. Finally, a pair of OH radicals form on the surface.

In addition, we introduce another water molecule to further study the interaction between water and the interfacial oxygen on 1 ML MgO(100)/Ag(100) surface. At the beginning of interaction, the second $H_2O$ molecule tends to approach the previously formed pair of OH groups by forming one hydrogen bond (see Figure 8(a)), and the corresponding adsorption energy is -0.91 eV per $H_2O$. In the following, the remaining interfacial O atom is also pulled out, and then two more OH groups form on the surface by the interaction of water molecule with the O atom. Interestingly, we find that OH groups are likely to form a chain of OH, as shown in Figure 8(b), and it is worth mentioning that there is remarkable structural distortion for surface Mg. In this case, the total energy is further reduced by 0.94 eV compared with that shown in Figure 8(a). Finally, two pairs of OH groups form by the interaction of $O_2$ with two water molecules on 1 ML MgO(100)/Ag(100) surface.

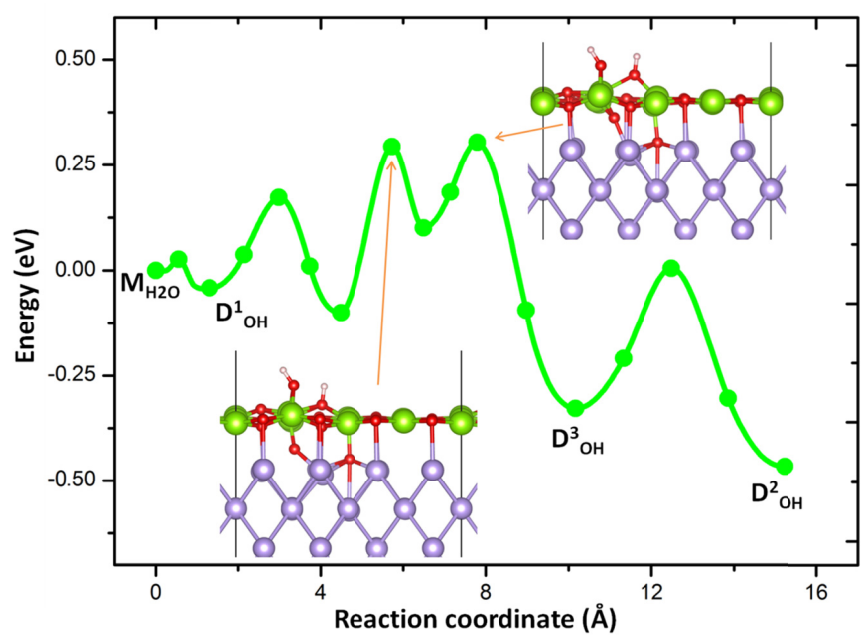

**Figure 7.** The possible potential energy profile for the interaction of $H_2O$ with interfacial O atoms to form OH groups on 1 ML MgO(100)/Ag(100) surface.

From the differential charge density distribution (Figure 8(c)), we further confirm that all the OH groups and the interfacial O gain electrons from the nearby atoms. The Bader charge analysis shows that OH and the interfacial O gain 0.84 e and 1.12 e, respectively, suggesting strong chemical bonds between adsorbates and the surface. Water molecules prefer to interact strongly with the dissociated $O_2$ to form chains of OH groups. From the differential charge density shown in Figure 8(d), OH groups tend to withdraw electrons (yellow regions) from 1 ML MgO(100)/Ag(100) surface. From the calculated Bader charges, we find that each OH group gain 0.67 e from the surface Mg atoms, which also indicate the formation of hydroxyl group on the surface.

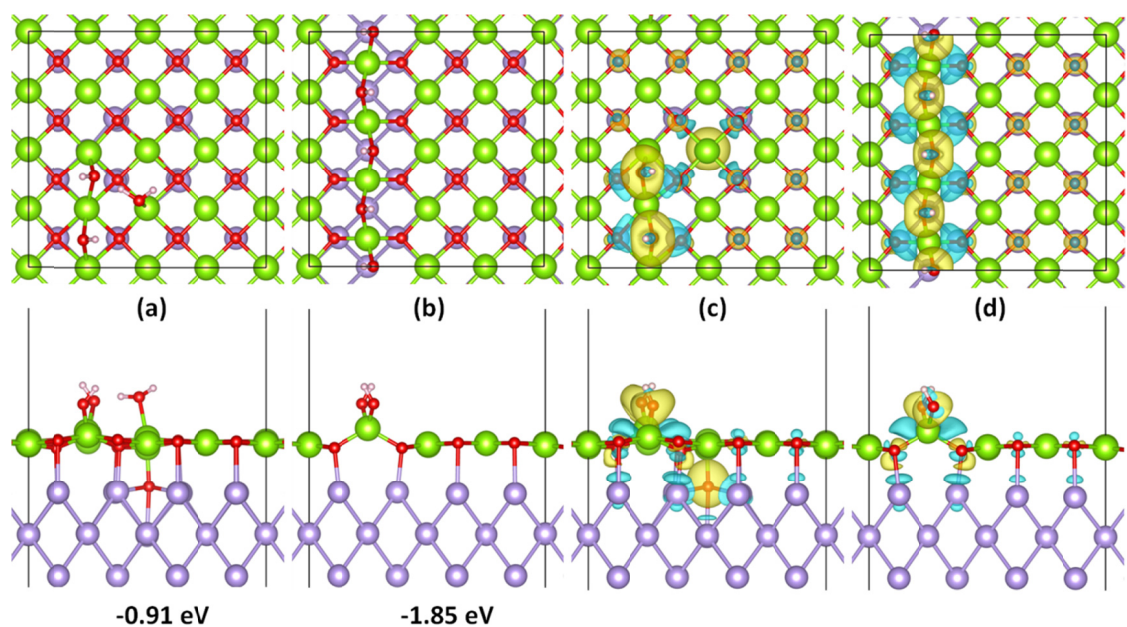

**Figure 8.** The adsorption of water molecule in (a) molecular and (b) dissociative forms on the 1 ML MgO(100)/Ag(100) surface with the pre-adsorbed water in the dissociative form. Top and side views of differential charge densities for (c) one water and (d) two water molecules in dissociative forms on the 1 ML MgO(100)/Ag(100) surface in the dissociative form. The used isosurface values are 0.004 e/bohr$^3$ (yellow: charge accumulation, ice-blue: charge depletion).

## ■CONCLUSION

In summary, we have performed first-principles calculations to study the enhanced surface reactivity by $O_2$ activation on 1 ML MgO(100)/Ag(100) surface. The adsorption configurations, reaction pathways, energy barriers, molecular dynamics simulations, and electronic properties are reported. Our results show that the adsorbed $O_2$ prefers to dissociate and diffuse into the interfacial region by forming reactive oxygen atoms. Interestingly, we find that water molecules act as pumps that is able to extract the interfacial oxygen, and as a result hydroxyl groups can be formed. The reaction mechanism for $H_2O$ and $O_2$ on 1 ML MgO(100)/Ag(100) surface can be described as follows:

$$O_2 \rightarrow 2O_i \quad (1)$$
$$2H_2O + 2O_i \rightarrow 4OH \quad (2)$$

This reaction is sustainable that 1 ML MgO(100)/Ag(100) surface plays an important role as a good catalyst. More importantly, the dissociated $O_2$ on 1 ML MgO(100)/Ag(100) may prompt more surface catalytic reactions, especially those involving water and oxygen.

## ■AUTHOR INFORMATION

Corresponding Author
*E-mail: xu.h@sustc.edu.cn
**Notes**
The authors declare no competing financial interest.

■ACKNOWLEDGMENTS

This work was supported by the National Natural Science Foundation of China (Grant Nos.11204185, 11334003, and 11404159).

■REFERENCES

(1) Linsebigler, A. L.; Lu, G.; Yates, J. T., Photocatalysis on TiO2 Surfaces: Principles, Mechanisms, and Selected Results. *Chem. Rev.* **1995**, 95, 735–758.

(2) Henderson, M. A., A Surface Science Perspective on TiO2 Photocatalysis. *Surf. Sci. Rep.* **2011**, 66, 185–297.

(3) Dohnálek, Z.; Lyubinetsky, I.; Rousseau, R., Thermally-driven Processes on Rutile TiO2(110)-(1×1): A Direct View at Atomic Scale. *Prog. Surf. Sci.* **2010**, 85, 161–205.

(4) Diebold, U., The Surface Science of Titanium Dioxide. *Surf. Sci. Rep.* **2003**, 48, 53–229.

(5) Cui, Y.; Shao, X.; Baldofski, M.; Sauer, J.; Nilius, N.; Freund, H., Adsorption, activation, and dissociation of oxygen on doped oxides. *Angew. Chem. Int. Ed.* **2013**, 52, 11385-11387.

(6) Di Valentin, C.; Ferullo, R.; Binda, R.; Pacchioni, G., Oxygen vacancies and peroxo groups on regular and low-coordinated sites of MgO, CaO, SrO, and BaO surfaces. *Surface Science* **2006**, 600, 1147-1154.

(7) Shin, H.-J.; Jung, J.; Motobayashi, K.; Yanagisawa, S.; Morikawa, Y.; Kim, Y.; Kawai, M., State-selective dissociation of a single water molecule on an ultrathin MgO film. *Nat. Mater.* **2010**, 9, 442–447.

(8) Jung, J.; Shin, H.; Kim, Y.; Kawai, M., Activation of ultrathin oxide films for chemical reaction by interface defects. *J. Am. Chem. Soc.* **2011**, 133, 6142.

(9) Jung, J.; Shin, H.; Kim, Y.; Kawai, M., Controlling water dissociation on an ultrathin MgO film by tuning film thickness. *Phys, Rev. B* **2010**, 82, 085413.

(10) Jung, J.; Shin, H.; Kim, Y.; Kawai, M., Ligand field effect at oxide-metal interface on the chemical reactivity of ultrathin oxide film surface. *J. Am. Chem. Soc.* **2012**, 134, 10554.

(11) Honkala, K.; Hellman, A.; Gronbeck, H., Water dissociation on MgO/Ag(100): Support induced stabilization or electron pairing? *J. Phys. Chem. C* **2010**, 114, 7070.

(12) Savio, L.; Celasco, E.; Vattuone, L.; Rocca, M. Enhanced reactivity at metal-oxide interface: Water interaction with MgO ultrathin films. *J. Phys. Chem. B* **2004**, 108, 7771.

(13) Hellman, A.; Klacar, S.; Gronbeck, H., Low Temperature CO Oxidation over Supported Ultrathin MgO Films. *J. Am. Chem. Soc.* **2009**, 131, 16636-16637.

(14) Gonchar, A.; Risse, T.; Freund, H. J.; Giordano, L.; Di Valentin, C.; Pacchioni, G., Activation of Oxygen on MgO: O-2(Center Dot-) Radical Ion Formation on Thin, Metal-Supported MgO(001) Films. *Angew. Chem. Int. Ed.* **2011**, 50, 2635-2638.

(15) Shin, D.; Sinthika, S.; Choi, M.; Thapa, R.; Park, N., Ab Initio Study of Thin Oxide-Metal over Layers as an Inverse Catalytic System for Dioxygen Reduction and Enhanced Co Tolerance. *Acs Catal* **2014**, 4, 4074-4080.

(16) Sun, Y. N.; Giordano, L.; Goniakowski, J.; Lewandowski, M.; Qin, Z. H.; Noguera, C.; Shaikhutdinov, S.; Pacchioni, G.; Freund, H. J., The Interplay between Structure and Co Oxidation Catalysis on Metal-Supported Ultrathin Oxide Films. *Angew. Chem. Int. Ed.* **2010**, 49, 4418-4421.

(17) Pal, J.; Smerieri, M.; Celasco, E.; Savio, L.; Vattuone, L.; Ferrando, R.; Tosoni, S.; Giordano, L.; Pacchioni, G.; Rocca, M., How Growing Conditions and Interfacial Oxygen Affect the Final Morphology of MgO/Ag(100) Films. *J. Phys. Chem. C* **2014**, 118, 26091.

(18) Smerieri, M.; Pal, J.; Savio, L.; Vattuone, L.; Ferrando, R.; Tosoni, S.; Giordano, L.; Pacchioni, G.; Rocca, M., Spontaneous Oxidation of Ni Nanoclusters on MgO Monolayers Induced by Segregation of Interfacial Oxygen. *J. Phys. Chem. Lett.* **2015**, 6, 3104.

(19) Kresse, G.; Hafner, J., Abinitio Molecular-Dynamics for Liquid-Metals. *Phys. Rev. B* **1993**, 47, 558-561.

(20) Kresse, G.; Hafner, J., Ab-Initio Molecular-Dynamics for Open-Shell Transition-Metals. *Phys. Rev. B* **1993**, 48, 13115-13118.

(21) Kresse, G.; Hafner, J., Ab-Initio Molecular-Dynamics Simulation of the Liquid-Metal Amorphous-Semiconductor Transition in Germanium. *Phys. Rev. B* **1994**, 49, 14251-14269.

(22) Kresse, G.; Joubert, D., From Ultrasoft Pseudopotentials to the Projector Augmented-Wave Method. *Phys. Rev. B* **1999**, 59, 1758-1775.

(23) Blochl, P. E., Projector Augmented-Wave Method. *Phys Rev B* **1994**, 50, 17953-17979.

(24) Hafner, J., Ab-Initio Simulations of Materials Using Vasp: Density-Functional Theory and Beyond. *J. Comput. Chem.* **2008**, 29, 2044-2078.

(25) Perdew, J. P.; Burke, K.; Ernzerhof, M., Generalized Gradient Approximation Made Simple. *Phys. Rev. Lett.* **1996**, 77, 3865-3868.

(26) Perdew, J. P.; Burke, K.; Ernzerhof, M., Generalized Gradient Approximation Made Simple (Vol 77, Pg 3865, 1996). *Phys. Rev. Lett.* **1997**, 78, 1396-1396.

(27) Henkelman, G.; Uberuaga, B. P.; Jonsson, H., A Climbing Image Nudged Elastic Band Method for Finding Saddle Points and Minimum Energy Paths. *J Chem. Phys.* **2000**, 113, 9901-9904.

(28) Bader, R. F. W, A quantum-theory of molecular-structure and its applications. *Chem. Rev.* **1991**, 91, 893.

(29) As implemented in the ABINIT code by K. Casek, F. Finocchi, and X. Conze.

(30) Alducin, M.; Busnengo, H.F.; Diez Muino, R., Dissociative Dynamics of Spin-triplet and Spin-singlet O2 on Ag(100). *J. Chem. Phys.* **2008**, 129, 224702.

(31) Frondelius, P.; Hellman, A.; Honkala, K.; Hakkinen, H.; Gronbeck, H., Charging of Atoms, Clusters, and Molecules on Metal-supported Oxides: A General and Long-ranged phenomenon. *Phys, Rev. B* **2008**, 78, 085426.

(32) Giordano, L.; Cinquini, F.; Pacchioni, G., Tuning the Surface Metal Work Function by Deposition of Ultrathin Oxide Films: Density Functional Calculations. *Phys. Rev. B* **2005**, 73, 045414.

(33) Cho, J.; Park, J. M.; Kim, K. S., Influence of intermolecular hydrogen bonding on water dissociation at the MgO(001) surface. *Phys. Rev. B* **2000**, 62, 9981.

(34) Song, Z.; Fan, J.; Xu. H, Strain-induced water dissociation on supported ultrathin oxide films. *Sci. Rep.* **2016**, 6, 22853.